\documentstyle[12pt]{article}

\newcommand{\EQ}{\begin{equation}}
\newcommand{\EN}{\end{equation}}
\newcommand{\bea}{\begin{eqnarray}}
\newcommand{\ena}{\end{eqnarray}}
\newcommand{\bdis}{\begin{displaymath}}
\newcommand{\edis}{\end{displaymath}}
\newcommand{\vs}[1]{\vspace{#1 mm}}

\renewcommand{\d}{\delta}
\renewcommand{\v}{\Delta}

\renewcommand{\t}{\tau}
\newcommand{\ep}{\epsilon}

\newcommand{\pa}{\partial}

\newcommand{\nn}{\nonumber \\}

\begin{document}

\topmargin 0pt
\oddsidemargin - 3mm

\newcommand{\NP}[1]{Nucl.\ Phys.\ {\bf #1}}
\newcommand{\PL}[1]{Phys.\ Lett.\ {\bf #1}}
\newcommand{\CMP}[1]{Comm.\ Math.\ Phys.\ {\bf #1}}
\newcommand{\PR}[1]{Phys.\ Rev.\ {\bf #1}}
\newcommand{\PRL}[1]{Phys.\ Rev.\ Lett.\ {\bf #1}}
\newcommand{\PREP}[1]{Phys.\ Rep.\ {\bf #1}}
\newcommand{\PTP}[1]{Prog.\ Theor.\ Phys.\ {\bf #1}}
\newcommand{\PTPS}[1]{Prog.\ Theor.\ Phys.\ Suppl.\ {\bf #1}}
\newcommand{\NC}[1]{Nuovo.\ Cim.\ {\bf #1}}
\newcommand{\JPSJ}[1]{J.\ Phys.\ Soc.\ Japan\ {\bf #1}}
\newcommand{\MPL}[1]{Mod.\ Phys.\ Lett.\ {\bf #1}}
\newcommand{\IJMP}[1]{Int.\ Jour.\ Mod.\ Phys.\ {\bf #1}}
\newcommand{\AP}[1]{Ann.\ Phys.\ {\bf #1}}
\newcommand{\RMP}[1]{Rev.\ Mod.\ Phys.\ {\bf #1}}
\newcommand{\PMI}[1]{Publ.\ Math.\ IHES\ {\bf #1}}
\newcommand{\JETP}[1]{Sov.\ Phys.\ J.E.T.P.\ {\bf #1}}
\newcommand{\TOP}[1]{Topology\ {\bf #1}}
\newcommand{\AM}[1]{Ann.\ Math.\ {\bf #1}}
\newcommand{\LMP}[1]{Lett.\ Math.\ Phys.\ {\bf #1}}
\newcommand{\CRASP}[1]{C.R.\ Acad.\ Sci.\ Paris\ {\bf #1}}
\newcommand{\JDG}[1]{J.\ Diff.\ Geom.\ {\bf #1}}
\newcommand{\JSP}[1]{J.\ Stat.\ Phys.\ {\bf #1}}

\begin{titlepage}
\setcounter{page}{0}
\begin{flushright}
January 1999\\
hep-th/9902001\\
\end{flushright}

\vs{5}
\begin{center}
{\Large  Algebraic Structure in 
   $0 < c \le 1$ \\ Open-Closed String Field Theories}

\vs{10}

{\large Daiji\ Ennyu\footnote{e-mail address:
ennyu@hiroshima-cmt.ac.jp}} \\
{\em 
Hiroshima National College of Maritime Technology \\ 
Toyota-Gun, Hiroshima 725-0200, Japan} \\
{\large Hiroshi\ Kawabe\footnote{e-mail address:
kawabe@yonago-k.ac.jp}} \\
{\em
Yonago National College of Technology \\ 
Yonago 683-8502, Japan} \\
and \\ 
{\large Naohito\ Nakazawa\footnote{e-mail address:
nakazawa@ifse1.riko.shimane-u.ac.jp}} \\
{\em 
Department of Physics, Faculty of Science and Engineering, Shimane University\\
Matsue 690-8504, Japan}  \\
\end{center}

\vs{8}
\centerline{{\bf{Abstract}}}

We apply stochastic quantization method to
Kostov's matrix-vector models for the second quantization of
orientable strings with Chan-Paton like factors, including both open and closed strings.
The Fokker-Planck hamiltonian deduces an orientable open-closed string field theory at the double scaling limit. 
There appears an algebraic structure in the continuum F-P hamiltonian including a Virasoro algebra and a $SU( r )$ current algebra.

\end{titlepage}
\newpage
\renewcommand{\thefootnote}{\arabic{footnote}}
\setcounter{footnote}{0}

The explicit construction of non-critical string field theories via the double scaling limit of the matrix models~\cite{DS} provides not only the basis for the non-perturbative analysis of string theories but also the clear understanding of the origin of 
the constraints realized in the algebraic structure of the string field theoretic hamiltonian~[1-10].  
The algebraic structure in non-critical string field theories has been investigated in this context for orientable closed strings~\cite{IK,JR,IIKMNS}, non-orientable closed strings~\cite{Na}, orientable open-closed strings~\cite{Kos,Mo} and non-orientable
 open-closed strings\cite{NE}.  
For orientable 2D surfaces with boundaries, an interesting algebraic structure, $sl (r, {\bf C})\times sl (r, {\bf C})$ chiral current algebra including $SU(r)$ current algebra, has been found in a field theoretic hamiltonian of orientable open-closed str
ings\cite{AJ}. 
The observation indicates the relation between the algebraic structure and a Chan-Paton like realization of gauge groups in open string theories. While the $SO(r)$ current algebra has been found in the string field theoretic hamiltonian for non-orientable
 open-closed strings\cite{NE}.

In this note, we construct non-critical orientable open-closed string field theories for $0 < c \le 1$ by applying stochastic quantization method to Kostov's matrix-vector model\cite{Kos,KK}. 
To introduce Chan-Paton like factors, we slightly modify Kostov's model. The continuum limit of the string field theory is taken at the double scaling limit of this matrix-vector model. 
We introduce the scale parameters on the boundaries which characterize the scaling behaviour of the time evolution along the boundaries. 
We show that the string field theoretic hamiltonian, which is the continuum limit of the loop space Fokker-Planck hamiltonian in stochastic quantization\cite{JR,Na}, appears in the form of a linear combination of three  deformation generators of orientabl
e strings which satisfy algebraic relations including a Virasoro algebra and a $SU( r )$ current algebra. 

For orientable 2D surfaces with boundaries, the sum of triangulated surfaces is  reconstructed by the hermitian matrix-vector models\cite{KK,Kaz}. 
To introduce the orientable open-closed string interactions for $0 < c \le 1$, we consider Kostov's matrix-vector model with Chan-Paton like factors. The action is given by\cite{Kos}  
\bea
\label{eq:action}
S_1 
&=& {\rm Tr} \big( 
{1\over2} \sum_{x,x'}A_{xx'}A_{x'x} - {1\over 3} {g\over \sqrt{N}}\sum_{x,x',x''}A_{xx'} A_{x'x''} A_{x''x}    
\big)         \ ,   \nn
S_2 
&=& \sum_{x,x'}\sum_{a=1}^r V^{a*}_x \big( 
\delta_{xx'} - {g^a_B \over\sqrt{N}} A_{xx'}
\big) V^a_{x'}        \ ,   
\ena
where the interactions of matrices, $A_{xx'}$, take place only for 
$ | x-x' | \le 1$ in the target space ${\bf Z}$, i.e. $x, x' \in {\bf Z}$. Namely, the relevant matrices are  $A_{xx\ ij}$, which are $N \times N$  hermitian matrices, and 
$A_{xx+1\ ij} = A^{\dagger}_{x+1x\ ij}$. 
$V^a_{x\ i}$ and $V^{a*}_{x\ i}$ are $N$-dimensional complex vectors with a Chan-Paton like factor $a$ which runs from 1 to $r$. 
The partition function, $Z$, and the free energy, $F$, of this model are given by  
\bea
\label{eq:partition-fun}
Z 
&\equiv& \int dA dV dV^* {\rm e}^{- S_1 - S_2}     \ ,  \nn
F
&\equiv& \ln Z         \nn
&=& \prod^r_{a=1}(g^a_B)^{L^a_B} g^S N^\chi     \ ,
\ena
where $S$ and $L^a_B$ are the total area and the total length of the boundary with  the \lq\lq colour\rq\rq index $a$ of the 2D surface, respectively. 
$\chi$ is the Eular number of the triangulated 2D surface with boundaries, 
$\chi \equiv 2 - 2 {\#}(handles) - {\#}(boundaries)$. 
By integrating out all the non-hermitian matrices, $A_{xx+1\ ij}$ and 
$A^{\dagger}_{xx+1\ ji}$, we obtain an effective action as a function of the 
hermitian matrix $M_{x\ ij} \equiv A_{xx\ ij} - {\sqrt{N}\over 2g}\d_{ij}$, 
and the vectors, 
$V^a_{x\ i}$ and $V^{a*}_{x\ i}$. We denote the effective action as 
$S_{\rm eff}(M, V, V^* )$ \cite{Kos}.

We define the time evolution, 
$
M_{x\ ij} ( \t+\v\t ) \equiv M_{x\ ij} ( \t ) + \v M_{x\ ij}( \t ) \ 
$
and 
$
V^a_{x\ i}( \t+\v\t ) \equiv V^a_{x\ i}( \t ) + \v V^a_{x\ i}( \t ) \ 
$, 
in terms of Ito's calculus\cite{I} with respect to the stochastic time $\t$ which is discretized with the unit time step $\v\t$. 
The time evolution is given by the following Langevin equations, 
\bea
\label{eq:Langevin}
{\v}M_{x\ ij}(\t)
&=&  - {\pa \over\pa M_{x\ ji}}S_{\rm eff}(M, V, V^* )(\t) \v\t + \v\xi_{x\ ij}(\t)        \ , \nn
{\v}V^a_{x\ i} (\t)
&=& - \lambda^a_x{\pa\over\pa V^{a*}_{x\ i}}S_{\rm eff}(M, V, V^* )(\t) \v\t 
+ \v\eta^a_{x\ i}(\t)   \ , \nn  
{\v}V^{a*}_{x\ i} (\t)
&=& - \lambda^a_x{\pa\over\pa V^a_{x\ i}}S_{\rm eff}(M, V, V^* )(\t) \v\t 
+ \v\eta^{a*}_{x\ i}(\t)    \ .  
\ena
We have introduced the scale parameter $\lambda^a_x$ in the Langevin equation of vector variables, $V^a_{x\ i}$ and $V^{a*}_{x\ i}$. 
At the double scaling limit, the parameter $\lambda^a_x$ defines the time scale for the stochastic time evolution of the open string end-point along the boundaries with the \lq\lq colour\rq\rq index $a$. 

The correlations of the white noise $\v\xi_{x\ ij}$ and $\v\eta^a_{x\ i}$ are 
defined by
\bea
\label{eq:noise}
<\v\xi_{x\ ij}(\t) \v\xi_{x'\ kl}(\t)>_\xi
&=& 2\v\t \d_{il} \d_{jk} \d_{xx'}  \ , \nn
<\v\eta^{a*}_{x\ i}(\t) \v\eta^b_{x'\ j}(\t)>_\eta
&=& 2\lambda^a_x \v\t \d^{ab}\d_{ij} \d_{xx'}            \ .
\ena
We will see later that the $\lambda^a_x$-independence of the $equilibrium$ is realized as the algebraic relation between the constraints in the string field theoretic hamiltonian which is equivalent to the integrability condition of the multi-time evoluti
on of the Langevin equations.

Now we define the closed strings,
$
\phi_x (L) =  N^{-1}{\rm tr}({\rm e}^{L N^{- 1/2} M_x})  \ , 
$
and the open strings,
$
\psi_x^{ab} (L) = N^{- 1 }( V^{a*}{\rm e}^{L N^{- 1/2} M_x} V^b )  \ . 
$
Following to Ito's calculus, we calculate the time development of these string variables\cite{Na},
\bea
\label{eq:loop-Langevin}
\v\phi_x (L)
&=& \v\t L \big\{  \int^L_0\! d\!L' \phi_x ( L') \phi_x ( L - L')    
+ \sum_{x'}\int^\infty_0\! d\!L' C^{(c)}_{x x'}\phi_x (L + L') \phi_{x'} (L')
   \nn 
&+& {1\over N g}\sum_{a, b, x'}g^a_B g^b_B 
\int^\infty_0\! d\!L' L' C^{(o)}_{x x'}\psi^{ab}_x (L + L') \psi^{ba}_{x'} (L')              \nn 
&+& {1\over N}\sum_{a = 1}^r g^a_B \psi^{aa}_x (L)  
  - {1\over 4g} \phi_x (L) + g {\pa^2 \over \pa L^2}\phi_x (L)  \big\}
 +   \v \zeta_x (L)   \ ,   \nn
\v\psi^{ab}_x (L)
&=& \lambda^a_x \v\t \big\{ 
 ( - 1 + {g^a_B\over 2g} + g^a_B {\pa\over \pa L} )\psi^{ab}_x (L)  
+ {1\over g}\sum_{c, x'}g^a_B g^c_B 
\int^\infty_0\! d\!L' C^{(o)}_{x x'}\psi^{cb}_x (L + L') \psi^{ac}_{x'} (L')  \big\} \nn 
&+& \lambda^b_x \v\t \big\{ 
 ( - 1 + {g^b_B\over 2g} + g^b_B {\pa\over \pa L} )\psi^{ab}_x (L)  
+ {1\over g}\sum_{c, x'}g^b_B g^c_B 
\int^\infty_0\! d\!L' C^{(o)}_{x x'}\psi^{ac}_x (L + L') \psi^{cb}_{x'} (L')  \big\} \nn
&+&   2\v\t \lambda^a_x \d^{ab}\phi_x (L)               \nn
&+& \v\t \big\{ 
  - {L\over 4g} \psi^{ab}_x (L) + Lg {\pa^2 \over \pa L^2}\psi^{ab}_x (L)   
+ L \sum_{x'}\int^\infty_0\! d\!L' C^{(c)}_{x x'}\psi^{ab}_x (L + L') \phi_{x'} ( L') \nn 
&+& \sum_c g^c_B \int^L_0\! d\!L'
   \psi^{ac}_x (L') \psi^{cb}_x  (L - L')     \nn
&+& \sum_{c,d, x'}{g^c_B g^d_B \over g}\!
\int^L_0\!\! d\!L'\!\!\int^\infty_0\!\! d\!L''\!\! \int^\infty_0 \!\!d\!L'''  
C^{(o)}_{x x'}\psi^{ad}_x (L' \!+\! L'') \psi^{cb}_x (L \!-\! L'\! + \! L''' )
 \psi^{dc}_{x'} (L'' \!+\! L''')                  \nn
&+& 2 \int^L_0\! d\!L' L' \psi^{ab}_x (L') \phi_x ( L-L')   \big\}       
+ \v\zeta^{ab}_x (L)   \ . 
\ena
Here we have introduced the adjacency matrices, 
$C^{(c)}_{xx'} = C^{(o)}_{xx'} = \d_{xx'+1} + \d_{xx'-1}$.  
In (\ref{eq:loop-Langevin}), the new noise variables are defined by
\bea
\label{eq:def-noise}
\v\zeta_x (L) 
&\equiv& L N^{-3/2}{\rm tr}({\rm e}^{L N^{- 1/2} M_x} \v\xi_x )    \ , \nn 
\v\zeta^{ab}_x (L) 
&\equiv&   N^{-3/2}\int^L_0\! d\!L' V^{a*}_x {\rm e}^{L' N^{- 1/2} M_x} \v\xi_x 
       {\rm e}^{(L-L') N^{- 1/2} M_x} V^b_x       \nn 
&+& N^{-1} \big\{ V^{a*}_x{\rm e}^{L N^{- 1/2} M_x} \v\eta^b_x 
       + \v\eta^{a*}_x{\rm e}^{L N^{- 1/2} M_x}V^b_x \big \}               \ .
\ena
%
%
%
We notice that 
$
<\v\zeta_x (L)>_{\xi\eta} = <\v\zeta^{ab}_x (L)>_{\xi\eta} 
 = 0   
$
 hold in the sense of Ito's calculus. 
Especially, from eqs. (\ref{eq:loop-Langevin}), the $\lambda$-independence of the equilibrium limit deduces a S-D equation which ensures the $SU( r )$ current algebra.

We can describe a class of matrix-vector models 
for the central charge $0 < c \le 1$ with the following adjacency matrices\cite{Kos}.
\bea
C^{(c)}_{xx'} 
&=& \cos (\pi p_0 ) \big( \d_{xx'+1} + \d_{xx'-1}   \big)    \ , \nn
C^{(o)}_{xx'} 
&=& \cos ( \pi p_0 /2 ) \big( \d_{xx'+1} + \d_{xx'-1}  \big)  \ .
\ena
The case, $p_0 = 1/m$, corresponds to the central charge, 
$c = 1 - {6\over m (m+1)}$.

The stochastic process is interpreted as the time evolution in a string field theory. 
The corresponding non-critical orientable open-closed string field theory is defined by the F-P hamiltonian operator. 
In terms of the expectation value of an observable $O(\phi, \psi)$, a
function of $\phi_x (L)$'s and $\psi^{ab}_x (L)$'s, 
the F-P hamiltonian operator ${\hat H}_{FP}$ is defined by\cite{Na}
\EQ
\label{eq:def-H}
<\phi (0), \psi (0)| {\rm e}^{- \t {\hat H}_{FP} } O({\hat \phi}, {\hat \psi})|0>
\equiv <O( \phi_{\xi\eta}(\t), \psi_{\xi\eta}(\t) )>_{\xi\eta}                  \  .
\EN
In R.H.S., $\phi_{\xi\eta}(\t)$ and $\psi_{\xi\eta}(\t)$ denote the solutions of the Langevin equations (\ref{eq:loop-Langevin}) 
with the appropriate initial configuration $\phi_x (L; \t =0)$ and $\psi^{ab}_x (L; \t =0)$. 
In L.H.S.,  
${\hat H}_{FP}$ is given by the differential operator in the well-known Fokker-Planck equation for the expectation value of the observable $O({\phi}, {\psi})$ 
by replacing the closed ( open ) string variable $\phi_x (L) $ ( $\psi^{ab}_x (L)$ ) to the creation operator 
${\hat \phi}_x (L)$ ( ${\hat \psi}^{ab}_x (L)$ ) and the differential ${\pa \over \pa \phi_x (L)}$ ( ${\pa \over \pa \psi^{ab}_x (L)}$ ) 
to the annihilation operator ${\hat \pi}_x (L)$ ( ${\hat \pi^{ab}_x (L)}$ ), respectively. 


The continuum limit of ${\hat H}_{FP}$ is taken by introducing a length 
scale
\lq\lq $\ep$ " which defines the physical length of the strings as 
$l \equiv L \ep$. 
At the double scaling limit $\ep \rightarrow 0$, we keep the string coupling constant,
$
G
\equiv N^{-2} \ep^{- 2D}       \ 
$, 
to be finite. 
The continuum stochastic time is given by  
$
d\t
\equiv \ep^{- 2 + D} \v\t       \ .
$
While we define the scaling of the scale parameter $\lambda^a_x$ by 
%
$
\lambda^a \equiv \ep^{- 1/2 - D/2 } \lambda^a_x       \ .
$
%
We also redefine field variables as follows.
\bea
\label{eq:renormalization}
\Phi_x (l)
&\equiv& \ep^{- D } {\hat \phi}_x (L)      \ , \nn
\Pi_x (l)
&\equiv& \ep^{- 1 + D } {\hat \pi}_x (L)      \ ,\nn
\Psi^{ab}_x (l)
&\equiv& \ep^{- 1/2 - D/2 } {\hat \psi}^{ab}_x (L)      \ , \nn
\Pi^{ab}_x (l)
&\equiv& \ep^{- 1/2 + D/2 } {\hat \pi}^{ab}_x (L)      \ .
\ena
The commutation relations are 
$
\big[ \Pi_x (l) ,  \Phi_{x'} (l')  \big]
= \d_{xx'}\d (l - l')               \ 
$
and 
$
\big[ \Pi^{ab}_x (l) ,  \Phi^{cd}_{x'} (l')  \big]
= \d^{ac}\d^{bd}\d_{xx'}\d (l - l')       \  
$
.
Then we obtain the continuum F-P hamiltonian, ${\cal H}_{FP}$, from
${\hat H}_{FP}$,
\bea
\label{eq:continuum-H}
{\cal H}_{FP}
&=& \sum_x  \int_0^{\infty}\!dl
l\big\{\sum_{x'}C_{xx'}\int_0^{\infty}\!dl' \Phi_x(l+l')\Phi_{x'}(l') 
+ \sqrt{G}\sum_{a,b,x'}C_{xx'}\int_0^{\infty}\!dl'l'\Psi_x^{ab}(l+l')\Psi_{x'}^{ba}(l')  \nn
&+& \int_0^l\!dl'\Phi_x(l')\Phi_x(l-l')  
+ G \int_0^{\infty}\!dl'l'\Phi_x(l+l')\Pi_x(l')
\big\}\Pi_x(l)    \nn
&+& \sum_{a,b,x} \int_0^{\infty}\!dl \big\{l\sum_{x'}C_{xx'} 
\int_0^{\infty}\!dl'\Psi_x^{ab}(l+l')\Phi_{x'}(l')    \nn
&+& \sum_{c,d,x'}C_{xx'} \int_0^l\!dl_1 \int_0^{\infty}\!dl_2 \int_0^{\infty}\!dl_3 \Psi_x^{ad}(l_1+l_2) \Psi_x^{cb}(l-l_1+l_3) \Psi_{x'}^{dc}(l_2+l_3)   \nn
&+& 2\int_0^l\!dl'l'\Psi_x^{ab}(l') \Phi_x(l-l')     \nn
&+& \lambda^a \sum_{c,x'}C_{xx'} \int_0^{\infty}\!dl' \Psi_x^{cb}(l+l') \Psi_{x'}^{ac}(l')  + \lambda^b \sum_{c,x'}C_{xx'} \int_0^{\infty}\!dl' \Psi_x^{ac}(l+l') \Psi_{x'}^{cb}(l') 
  \nn 
 &+& (\lambda^a +\lambda^b)\delta^{ab} \Phi_x(l)
\big\} \Pi_x^{ab}(l)      \nn
&+& \sum_{a,b,c,d,x} \sqrt{G} \int_0^{\infty}\!dl \int_0^{\infty}\!dl' \int_0^l \!dl_1 \int_0^{l'}\!dl'_1 \Psi_x^{ad}(l_1+l'_1) \Psi_x^{cb}(l+l'-l_1-l'_1) \Pi_x^{ab}(l) \Pi_x^{cd}(l')    \nn
&+& \sum_{a,b,c,x} \sqrt{G} \int_0^{\infty}\!dl \int_0^{\infty}\!dl' 
\big\{
\lambda^a \Psi_x^{cb}(l+l') \Pi_x^{ab}(l) \Pi_x^{ca}(l')
 + \lambda^b \Psi_x^{ac}(l+l') \Pi_x^{ab}(l) \Pi_x^{bc}(l')
\big\}     \nn
&+& 2G\sum_{a,b,x} \int_0^{\infty}\!dl \int_0^{\infty}\!dl'  ll'\Psi_x^{ab}(l+l') \Pi_x(l) \Pi_x^{ab}(l')              \  . 
\ena

The continuum F-P hamiltonian (\ref{eq:continuum-H}) takes the form of a linear combination of three continuum generators of string deformation, 
\bea
\label{eq:algebraic-H}
{\cal H}_{FP} 
&=& \sum_x \int_0^{\infty}\!dl l {\cal L}_x (l) \Pi_x(l)
+ \sum_{a,b,x} \int_0^{\infty}\!dl 
\big\{ \lambda^a {\cal J}_x^{ab}(l) +\lambda^b {\cal J}_x^{ba \ast}(l) \big\} \Pi_x^{ab}(l)  \nn
&+& \sum_{a,b,x} \int_0^{\infty}\!dl \Big\{ l {\cal K}_x^{ab}(l)
 - \sum_c \int_0^l\!dl' l' \Big( {\cal J}_x^{cb}(l') \Psi_x^{ac}(l-l')     \nn
  &{}& \quad +{\cal J}_x^{ca \ast}(l') \Psi_x^{bc \ast}(l-l') \Big) \Big\} \Pi_x^{ab}(l) \ .
\ena
The explicit forms of these generators are given by 
\bea
\label{eq:def-current}
{\cal L}_x(l)
&=& \sum_{x'}C_{xx'} \int_0^{\infty}\!dl' \Phi_x(l+l') \Phi_{x'}(l')      \nn
&+& \sqrt{G} \sum_{a,b,x'}C_{xx'} \int_0^{\infty}\!dl' l' \Psi_x^{ab}(l+l') \Psi_{x'}^{ba}(l')
+\int_0^l\!dl' \Phi_x(l') \Phi_x(l-l')       \nn
&+& G \int_0^{\infty}\!dl' l' \Phi_x(l+l') \Pi_x(l')
+G \sum_{a,b} \int_0^{\infty}\!dl' l' \Psi_x^{ab}(l+l') \Pi_x^{ab}(l')       \ , \nn
{\cal K}_x^{ab} (l) 
&=& \sum_{x'}C_{xx'} \int_0^{\infty}\!dl' \Psi_x^{ab}(l+l') \Phi_{x'}(l')    \nn
&+& G \int_0^{\infty}\!dl' l' \Psi_x^{ab}(l+l') \Pi_x(l')
+\sum_{c,d,x'}C_{xx'} \int_0^{\infty}\!dl_1 \int_0^{l+l_1}\!dl_2 \Psi_x^{ac}(l_2)           \Psi_x^{db}(l+l_1-l_2) \Psi_{x'}^{cd}(l_1)      \nn
&+& \sqrt{G} \sum_{c,d} \int_0^{\infty}\!dl_1 \int_0^{l+l_1}\!dl_2 \Psi_x^{ac}(l_2) \Psi_x^{db}(l+l_1-l_2) \Pi_{x}^{dc}(l_1)        \nn
&+& 2 \int_0^l\!dl' \Psi_x^{ab}(l') \Phi_x(l-l')     \ , \nn
{\cal J}_x^{ab} (l) 
&=& \delta^{ab} \Phi_x(l) 
+\sum_{c,x'} C_{xx'} \int_0^{\infty}\!dl' \Psi_x^{cb}(l+l') \Psi_{x'}^{ac}(l')  \nn 
&+& \sqrt{G} \sum_c \int_0^{\infty}\!dl' \Psi_x^{cb}(l+l') \Pi_x^{ca}(l')   \ . 
\ena

These generators satisfy the following algebra including the Virasoro algebra and the $SU( r )$ current algebra,
\bea
\label{eq:algebra}
\big[{\cal L}_x(l), {\cal L}_{x'}(l') \big]
&=& - (l-l') G \delta_{xx'} {\cal L}_{x} (l+l')      \ , \nn
\big[ {\cal L}_x (l),  {\cal K}_{x'}^{ab} (l') \big]  
&=& - (l - l') G \delta_{xx'} {\cal K}_x^{ab}(l+l')         \nn 
&+& G \delta_{xx'} \sum_c \int_0^l\!du (l-u) \big\{ {\cal J}_x^{cb}(l+l'-u) \Psi_x^{ac}(u) + {\cal J}_x^{ca \ast}(l+l'-u) \Psi_x^{bc \ast}(u) \big\} 
    \ , \nn
\big[ {\cal L}_x (l),  {\cal J}_{x'}^{ab} (l') \big] 
&=& l' G \delta_{xx'} {\cal J}_x^{ab}(l+l')      \ , \nn
\big[ {\cal J}_x^{ab} (l),  {\cal K}_{x'}^{cd}(l') \big] 
&=& \sqrt{G} \delta^{ad} \delta_{xx'} {\cal K}_x^{cb}(l+l')     \nn
&-& \sqrt{G} \delta^{ad} \delta_{xx'} \sum_e \int_0^l\!du {\cal J}_x^{ec*}(l+l'-u) \Psi_x^{be*}(u)       \nn
&-& \sqrt{G} \delta_{xx'} \int_0^l\!du {\cal J}_x^{ad}(l+l'-u) \Psi_x^{cb}(u) 
      \ , \nn 
\big[ {\cal J}_x^{ab} (l),  {\cal J}_{x'}^{cd} (l') \big] 
&=& \sqrt{G} \d^{ad} \d_{xx'} {\cal J}_x^{cb}(l+l') - \sqrt{G}\d^{cb}\d_{xx'} {\cal J}^{ad}(l+l')    \ , \nn
\big[ {\cal K}^{ab} (l),  {\cal K}^{cd} (l') \big] 
&=& \sqrt{G} \d_{xx'} \sum_e \int_0^{l'}\!du \int_0^{l'-u}\!dv 
\big\{ {\cal J}_x^{eb}(l+l'-u-v)\Psi_x^{ad}(u)\Psi_x^{ce}(v)        \nn 
&+& {\cal J}_x^{ea*}(l+l'-u-v)\Psi_x^{de*}(u)\Psi_x^{cb}(v)   \big\}     \nn
 &-& \sqrt{G} \d_{xx'} \sum_e \int_0^l\!du \int_0^{l-u}\!dv 
\big\{ {\cal J}_x^{ed}(l+l'-u-v)\Psi_x^{cb}(u)\Psi_x^{ae}(v)      \nn 
&+& {\cal J}_x^{ec*}(l+l'-u-v)\Psi_x^{ad}(u)\Psi_x^{be*}(v) \big\}
         \  .    
\ena
In the precise sense, these commutation relations are not an algebra because the open string creation operators $\Psi^{ab}(l)$ appear in the R.H.S. 
of (\ref{eq:algebra}). The algebraic structure (\ref{eq:algebra}) is  understood as the consistency condition for the constraints on the equilibrium expectation value or \lq\lq integrable condition\rq\rq  of the stochastic time evolution. 
The hermitian part of the current, ${\cal J}^{ab}(l) + {\cal J}^{*ba}(l) $, generates the S-D equation 
which implies the $\lambda$-independence of the equilibrium distribution. While the anti-hermitian part of the current, ${\cal J}^{ab}(l) - {\cal J}^{*ba}(l)$, satisfies $SU( r )$ current algebra.  
The similar algebraic structure is also found in Ref.\cite{AJ} in a discretized form without continuum limit.

In conclusion, we have derived the continuum 
F-P hamiltonian which defines the non-critical orientable  
open-closed string field theory for $0 < c \le 1$ by applying SQM to Kostov's 
matrix-vector model. 
The origin of the algebraic structure (\ref{eq:algebra}) can be traced to the noise correlations which generate the time evolution in the Langevin equation. 
Namely, the noise correlations realize the deformation of open and closed strings which are equivalent to those generated by the three constraints, 
${\cal L}(l)$, ${\cal J}^{ab}(l)$ and ${\cal K}^{ab}(l)$. 
We have shown that the structure to be universal representing the scale invariance in the non-critical orientable open-closed string theories.  
We hope that our approach is also usefull to the non-perturbative analysis of superstring theories such as type IIA, IIB and type I theories for which 
matrix models are proposed~\cite{IIA,IIB} by the large N reduction of super Yang-Mills theory.

N. N. would like to thank I. Kostov for valuable comments. 
This work was partially supported by the Minisitry of Education, 
Science and Culture, Grant-in-Aid for Exploratory Research,  
09874063, 1998.


\end{document}